\documentclass[10pt,aps,prb,superscriptaddress,twocolumn,a4paper,floatfix]{revtex4-2}

\renewcommand{\toprule}{\hline}
\newcommand{\midrule}{\hline}
\newcommand{\bottomrule}{\hline}

\usepackage{subfigure}
\usepackage{graphicx}
\usepackage{bm}
\usepackage{amsmath}
\usepackage{epstopdf}
\usepackage{dcolumn}
\usepackage[colorlinks=true, citecolor=blue, linkcolor=blue, urlcolor=blue]{hyperref}

\usepackage[dvipsnames]{xcolor} 
\usepackage{placeins} 
\usepackage{multirow}
\usepackage{inputenc}
\usepackage{braket}
\usepackage{soul} 

\newcommand{\betadd}{\beta_{dd}}
\newcommand{\betapd}{\beta_{pd}}

\newcommand{\Deg}{\Delta{}e_g}
\newcommand{\Dttwog}{\Delta{}t_{2g}}

\mathchardef\ordinarycolon\mathcode`\:
\mathcode`\:=\string"8000
\begingroup \catcode`\:=\active
  \gdef:{\mathrel{\mathop\ordinarycolon}}
\endgroup

\definecolor{annotcolor}{rgb}{.7,0,0}

\newcommand{\affilepfour}{Experimentelle Physik IV and R\"ontgen Research Center for Complex Materials (RCCM), Fakult\"at f\"ur Physik und Astronomie, Universit\"at W\"urzburg, Am Hubland, D-97074 W\"urzburg, Germany}

\newcommand{\affiltudk}{Department of Energy Conversion and Storage, Technical University of Denmark, DK-2800 Kgs. Lyngby, Denmark}
\newcommand{\affilusask}{Department of Physics and Engineering Physics, University of Saskatchewan, 116 Science Place, Saskatoon, Saskatchewan S7N 5E2, Canada}
\newcommand{\affilsbqmi}{Stewart Blusson Quantum Matter Institute, University of British Columbia, Vancouver, British Columbia, V6T 1Z4, Canada}
\newcommand{\affilcls}{Canadian Light Source, Saskatoon, Saskatchewan S7N 2V3, Canada}

\begin{document}

\title{Strain induced stabilization of a static Jahn--Teller distortion in the O$^*$-phase of La$_{7/8}$Sr$_{1/8}$MnO$_3$}

\author{M. Dettbarn}\affiliation{\affilepfour}
\author{V. B. Zabolotnyy}\affiliation{\affilepfour}
\author{A. Tcakaev}\affiliation{\affilepfour}
\author{R. Sutarto}\affiliation{\affilcls}
\author{F. He}\affiliation{\affilcls}
\author{Y. Z. Chen}\affiliation{\affiltudk}
\author{R. J. Green}\affiliation{\affilusask}\affiliation{\affilsbqmi}
\author{V. Hinkov}\email[]{hinkov@physik.uni-wuerzburg.de}\affiliation{\affilepfour}

\date{\today}

\begin{abstract}
At room temperature, bulk La$_{7/8}$Sr$_{1/8}$MnO$_3$ is in the dynamic Jahn--Teller O$^*$ phase, but undergoes a transition to a static, magnetically ordered Jahn--Teller phase at lower temperatures. Here we study a $6$ unit cells thin film of this compound grown on SrTiO$_3$, resulting in small compressive strain due to a lattice mismatch of $\lesssim 0.2\%$. We combine X-ray absorption spectroscopy with multiplet ligand field theory to study the local electronic and magnetic properties of Mn in the film. We determine the Mn $d_{3z^2-r^2}$ orbital to be $0.13\;\text{eV}$ lower in energy than the $d_{x^2-y^2}$, which is a disproportionately large splitting given the small degree of compressive strain. We interpret this as resulting from the strain providing a preferential orientation for the MnO$_6$ octahedra, which are strongly susceptible to such a deformation in the vicinity of the phase transition. Hence, they collectively elongate along the $c$ axis into a static Jahn--Teller arrangement. Furthermore, we demonstrate the strongly covalent character of La$_{7/8}$Sr$_{1/8}$MnO$_3$, with a contribution of nearly $50\%$ of the one-ligand-hole configuration $d^{5} \underline{L}^1$ to the ground state wavefunction. Finally, we find the system to be in a high-spin configuration, with the projection of the local magnetic moment on the quantization axis being about $3.7\;\mu_{\text{B}}/\text{Mn}$. We show, however, that the system is close to a high-spin--low-spin transition, which might be triggered by crystal field effects.
\end{abstract}

\keywords{XAS; crystal field calculations; manganites; manganates; lanthanum manganite; lanthanum strontium manganite; LMO; LSMO; transition metal compounds; transition metal oxides}
\maketitle

\section{Introduction}
\label{sec:Introduction}

Alkaline-earth lanthanum manganites, among which La$_{1-x}$Sr$_x$MnO$_3$ (LSMO) is the most widely studied representative, exhibit a variety of structural, magnetic and electronic phases \cite{Zhou_Goodenough_PhaseDiag, Zhou_Liu_Goodenough_PhaseDiag, Chmaissem_PhaseDiag, Chen_LSMO_COtr, Ramirez_CMR, Dagotto_Book}. These phases and the transitions between them are controlled by the interplay of temperature, doping and structural properties, which determine the interactions of the Mn $3d$ and O $2p$ electrons. In the bulk, LSMO exhibits a well-studied orthorhombic structure \cite{Zhou_Goodenough_PhaseDiag, Chmaissem_PhaseDiag, Zhou_Liu_Goodenough_PhaseDiag, Cox_LSMO, Uhlenbruck_LSMO} with a small tolerance factor and a strong Jahn--Teller effect typical for Mn$^{3+}$ systems \cite{Geck_OO, Huang_LMO, Dagotto_Book, Tang_simu}. It is ferromagnetic (FM) up to $x\sim0.5$ \cite{Zener_DEx, Anderson_Hasegawa_DEx, de_Gennes_DEx, Millis_1996_PRL, Urushibara_DEx}, except for an antiferromagnetic (AFM) window for $x\lesssim0.1$ \cite{Urushibara_DEx, Xiong_LSMO_JT, Dabrowski_PhaseDiag, Zhou_Goodenough_PhaseDiag, Chmaissem_PhaseDiag}.

In heterostructures substrate-induced strain and interface reconstructions can substantially modify the crystalline structure and alter the sequence of the energy levels found in the bulk. This can modify the local magnetic moments and the exchange interactions, and therefore the magnetic ordering. Indeed, emergent ferromagnetism has recently been observed in epitaxial LaMnO$_3$ (LMO) films \cite{Anahory, Renshaw-Wang, Niu}. Both for such LMO films, and for LSMO films with $x$ throughout the bulk FM window, ferromagnetism only occurs if their thickness $t$ exceeds $t_c \sim 4$\,u.\,c., suggesting a magnetically ``dead'' interfacial layer \cite{Niu, Kodama_Dead_Layer, Peng_Dead_Layer, Balcells_Dead_Layer, Borges_LSMO/MgO, Huijben_LSMO}. Ferromagnetic ordering is substantially affected not only by the film thickness, but also by strain \cite{Niu, Roqueta, Kim_0.12_LSMO, Hu_LSMO/LAO, Monsen_LSMO, Koohfar_2019, Koohfar_2020, Shibata_2018, Borges_LSMO/MgO, Huijben_LSMO}. Whereas Mn$^{3+}$ is omnipresent, Mn$^{4+}$ and the stoichiometrically unexpected Mn$^{2+}$ have been observed as well \cite{de_Jong, de_Jong_next, Niu, Kaspar, Wu_IF_induced, Chen_Accu_Eme, Quan_APL, Peng_Exchange_bias, Pellegrin_XMCD, Choi_LSMO_switching, Ning_Nanoscale}. Mn$^{3+}$ is a key ingredient of each FM theory in LSMO. Obviously, knowledge of its electronic structure and in particular of its local magnetic moment is invaluable. 

SQUID magnetometry is  a widely used method to probe Mn moments \cite{Niu, Jonker-van-Santen, Kim_0.12_LSMO, Kotani_Magnetic_bubbles, Hu_LSMO/LAO, Vaz_LSMO/PZT, Monsen_LSMO, Koohfar_2019, Koohfar_2020, Wahler_LSMO_nanostructures, Borges_LSMO/MgO, Huijben_LSMO, Curiale_Nanoparticles}. It typically measures the total magnetic moment of the sample. However, magnetometry is not valence specific, and sample inhomogeneities, in particular dead layers, complicate the estimation of the ordered volume fraction \cite{Kodama_Dead_Layer, Peng_Dead_Layer, Balcells_Dead_Layer, Borges_LSMO/MgO, Huijben_LSMO} and therefore of the individual Mn moments.

A method which is more direct, albeit at the expense of increased technical and analytic intricacy, is X-ray absorption spectroscopy (XAS). XAS spectra at the Mn-$L_{2,3}$ edge uniquely depend on the $3d$ level occupation and the Mn environment. By analyzing the spectra using multiplet ligand field theory (MLFT), one can understand this dependence on a microscopic level and obtain the individual magnetic moments undistorted by sample inhomogeneities.

A typical MLFT calculation relies on a set of microscopic parameters such as the strength and the symmetry of the crystal field, overlap integrals between involved atomic orbitals, including  the Slater integrals determining Coulomb interaction and the on-site electronic correlations \cite{Maurits_MLFT, Bocquet_1992_general, vdLK, Maurits_thesis}. In many cases, \textit{ab initio} values of these parameters already provide an adequate description of the underlying physics. Tuning them to enhance the agreement between calculated and experimental XAS spectra improves the numerical accuracy of the devised model and the conclusions based on it.

Earlier attempts to fit the spectral shape of Mn$^{3+}$ XAS using multiplet theory had some success \cite{Abbate_1992, Castleton/Altarelli, Taguchi/Altarelli, Kuepper, Cho_2009, Lee_Mo_SMO, Ghiasi, Cuartero}. A decent agreement was achieved by assuming a $D_{4h}$ symmetry, but the accuracy of the obtained MLFT model was not sufficient to draw a clear conclusion about the orbital polarizations or the degree of covalency \cite{Castleton/Altarelli, Cho_2009}.

In this work, using MLFT fits to XAS data, we investigate a La$_{1-x}$Sr$_x$MnO$_3$ ($x=1/8$) film with a thickness $t = 6$\,u.\,c., which is just above the critical thickness for ferromagnetism. In agreement with previous reports, the sample contains some amount of Mn$^{2+}$, whose contribution we subtract to obtain a pure Mn$^{3+}$ spectrum. We do not \emph{a priori} limit the fit to a few individual MLFT parameters, while fixing the others to empirical or \emph{ab initio} values. Rather we tune most of them within established physically meaningful intervals \cite{Saitoh_LXO_95, Bocquet_1992_general, Abbate_1992, Maurits_MLFT, de-Groot-Kotani_Book, Dagotto_Book}. This allows us to first separate out parameters to which our analysis is not sensitive. Then we focus our final fit on the crystal field splitting $10Dq$, the strain-induced splittings of the $e_g$ and $t_{2g}$ levels, $\Deg$ and $\Dttwog$, the charge transfer energy $\Delta$ and the intra-atomic relaxation scaling factors of the Slater integrals, $\beta_{dd}$ and $\beta_{pd}$.

We show that $D_{4h}$ symmetry describes our strained film well. Our fits yield a strongly covalent ground state with substantial contributions from both ionic $d^4\underline{L}^0$ and charge-transfer $d^5\underline{L}^1$ and $d^6\underline{L}^2$ configurations. Based on the fit results, we calculate the local Mn moment. Finally, we confirm the robustness of our fits by discussing two-dimensional fitness maps through the multidimensional fitting parameter space.

\section{Sample System and Experimental Technique}
\label{sec:Experiment}

The phase diagram of bulk La$_{1-x}$Sr$_x$MnO$_3$ is particularly complex between $x\approx0.1$ and $x\approx0.15$ \cite{Geck_OO_NJP, Kawano_LSMO, Cox_LSMO}. At temperatures well above room temperature, LSMO is in the O$^*$ phase. Whereas different authors report slightly different lattice constants, the consensus is that this phase is only weakly orthorhombic: According to Kawano \emph{et al.}, the O$^*$ phase is nearly cubic, the lattice parameters in orthorhombic notation being $a_{\textrm{o}}=5.537$\,$\textrm{\AA}$, $b_{\textrm{o}}=5.545$\,$\textrm{\AA}$ and $c_{\textrm{o}}/\sqrt{2}=5.530$\,$\textrm{\AA}$, hence $a_{\textrm{o}}\approx b_{\textrm{o}}\approx c_{\textrm{o}}/\sqrt{2}$, with a unit cell volume of $V_{\textrm{o}}=240.113$\,$\textrm{\AA}^3$ \cite{Kawano_LSMO}. The Mn--O distances are found to be nearly identical as well, which indicates the absence of a notable static Jahn--Teller distortion. However, a dynamic Jahn--Teller effect is assumed, with octahedral distortions unresolved by diffraction. Cox \emph{et al.} report a slightly larger spread of the three orthorhombic lattice constants ($a_{\textrm{o}}=5.5437$\,$\textrm{\AA}$, $b_{\textrm{o}}=5.5257$\,$\textrm{\AA}$ and $c_{\textrm{o}}/\sqrt{2}=5.5104$\,$\textrm{\AA}$, unit cell volume $V_{\textrm{o}}=238.719$\,$\textrm{\AA}^3$) \cite{Cox_LSMO}, as do Geck \emph{et al.} \cite{Geck_OO_NJP}. 

Close to room temperature, LSMO undergoes a transition to a statically Jahn--Teller ordered phase, O', followed by magnetic and charge-ordered phases at even lower temperatures. At $x=1/8$, the transition to the O' phase occurs around $280$\,K. Hence, we can assume that a bulk sample of our doping level would be in the O$^*$ phase and subject to dynamic Jahn--Teller fluctuations.

The popular substrate material SrTiO$_3$ (STO) has a cubic structure with $a_{\textrm{c}}=b_{\textrm{c}}=c_{\textrm{c}}=3.905$\,$\textrm{\AA}$ \cite{STO_structure_chapter} and a unit cell volume of $V_{\textrm{c}}=59.547$\,$\textrm{\AA}^3$, which corresponds to $V_{\textrm{o}}=238.190$\,$\textrm{\AA}^3$ in orthorhombic notation. To quantify the lattice mismatch of LSMO with the STO substrate, we use the relation $m=(a_{\textrm{l}}-a_{\textrm{s}})/a_{\textrm{s}}$, where $a_{\textrm{l}}$ and $a_{\textrm{s}}$ are the pseudocubic in-plane lattice constants of the film and the substrate, respectively. In order to apply the relation, we have calculated \cite{Vailionis} the pseudocubic lattice parameters of LSMO from the orthorhombic ones. 

For the Kawano results, we obtain $a_{\textrm{c}}=3.9103$\,$\textrm{\AA}$, $b_{\textrm{c}}=3.9181$\,$\textrm{\AA}$ (average: $3.9142$\,$\textrm{\AA}$) and $c_{\textrm{c}}=3.9181$\,$\textrm{\AA}$, which corresponds to a lattice mismatch of $0.24\%$, and thus a small compressive strain. Assuming pseudomorphic growth with unit cell volume conservation, the film out-of-plane lattice constant becomes $3.9366$\,$\textrm{\AA}$, i.e. $0.81\%$ larger than the in-plane lattice constant. For the Cox results, we obtain $a_{\textrm{c}}=3.8964$\,$\textrm{\AA}$, $b_{\textrm{c}}=3.9136$\,$\textrm{\AA}$ (average: $3.905$\,$\textrm{\AA}$) and $c_{\textrm{c}}=3.9136$\,$\textrm{\AA}$, which corresponds to zero mismatch ($0.0\%$) and strain, and a film out-of-plane lattice constant larger by $0.22\%$ than the in-plane one.

\begin{figure}[tb]
	\includegraphics[width = 1.0\columnwidth]{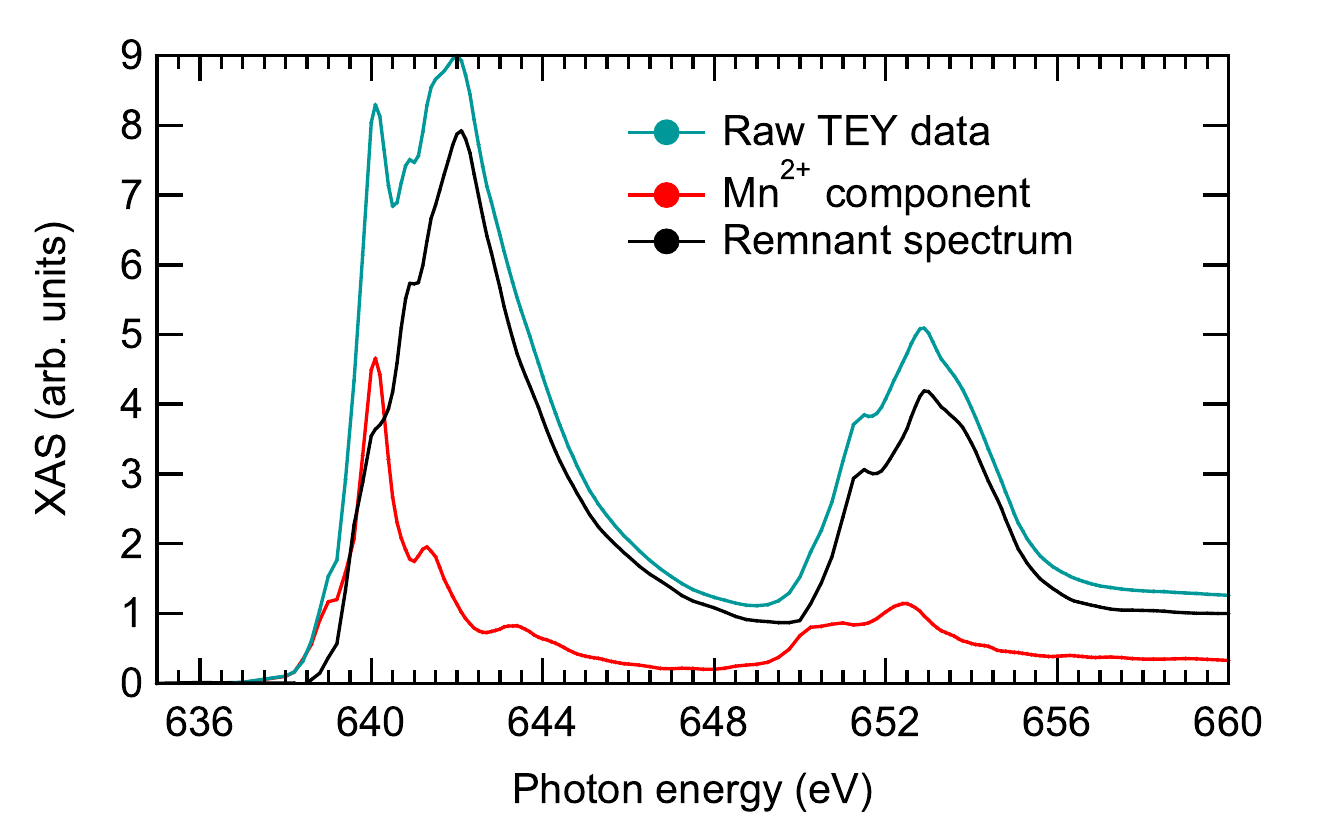}
	\caption{(Color online) Experimental X-ray absorption spectrum at the Mn $L_{2,3}$ edge, measured in TEY-mode (turquoise), shown together with subtracted reference Mn$^{2+}$ component (red) and the remnant spectrum (black). All spectra are normalized to the off-resonance theoretical scattering factor ($f_2$) as obtained from Chantler tables.}
	\label{fig:expcurve}
\end{figure}

The studied sample was a $6$ unit cells thin La$_{7/8}$Sr$_{1/8}$MnO$_3$ film epitaxially grown on an STO substrate using pulsed laser deposition (PLD), whose growth and characterization are described elsewhere \cite{Chen_LSMO_buffering}. An XAS spectrum of our sample was measured at the REIXS beamline \cite{REIXS} of the Canadian Light Source with $\sigma$-polarized light in total electron yield (TEY) mode at room temperature and is shown in Fig.~\ref{fig:expcurve}.

Owing to the thinness of the film, the saturation and self-absorption effects are negligible \cite{Thole_Lanthanides, vdL_TEY}, so the surface sensitive TEY signal can be considered a good approximation of the absorption coefficient $\sigma(E)$, which can be readily converted to the imaginary part of the scattering factor $f_2(E)$ \cite{de-Groot-Kotani_Book}. Since the Mn $L$-edge resonance spectrum does not overlap with resonances of any other elements present in the sample, their contribution can be modeled by a simple linear background.

We have corrected our data for the Mn$^{2+}$ contribution typical for this film thickness by subtracting a reference Mn$^{2+}$ XAS spectrum (see Fig.~\ref{fig:expcurve}). The reference spectrum has been obtained from a strongly reduced manganite film \cite{Chen_LSMO_buffering}. It has been subtracted from our measured data in such a way that the pre-$L_3$ edge at about $640\;\text{eV}$ does not become negative or exhibits a noticeable dip. As confirmed by our fits, the resulting spectrum contains no spectral features of Mn$^{4+}$ or other ions and represents pure Mn$^{3+}$. Indications for this already come from its close similarity to previously published absorption spectra of undoped LMO (nominal Mn$^{3+}$) \cite{Burnus_LMCO, Piamonteze_XMCD, Kitamura_Charge_transfer, Park/Cheong/Chen, Cho_2007} as well as LSMO compounds of varying doping \cite{Abbate_1992, Kavich_XMCD, Aruta_XMCD, Pellegrin_XMCD}.

\section{Theory}
\label{sec:Theory}

The off-resonant part of $f_2(E)$ consists of step edges due to electronic excitations from the $2p$ shell into the continuum. This contribution to the total $f_2(E)$ is readily available in Chantler tables \cite{Chantler}. Additionally we allow for a slight energy shift and smoothing of the step edges.

The resonant part of the Mn$^{3+}$ scattering factor $f_2(E)$ can be modeled using multiplet ligand field theory (MLFT), which explicitly includes ligand-to-metal hopping, in addition to crystal field splitting, Coulomb interaction and spin--orbit (SO) coupling. Most directly, an XAS spectrum can be written as a double sum over the Boltzmann-weighted dipole transitions from the initial states $|\psi_i\rangle$ into the final states $|\psi_f\rangle$:
\begin{equation}
S( \omega) \sim \frac{1}{Z} \sum\limits_{i, f} e^{-\frac{E_i}{kT}}  \left| \langle \psi_f | \hat{T}| \psi_i \rangle \right|^2 
 \delta (E_f - E_i - \hbar \omega),
\label{eq:CFT}
\end{equation}
where $\hat{T} = \frac{e}{m_e} \hat{\mathbf{p}} \cdot \mathbf{A}$ is the transition operator for the XAS spectroscopy, $\hat{\mathbf{p}}$ is the momentum operator of the electron, and $\mathbf{A}$ is the vector field of the photon. In practice, though, there is no need to calculate the full spectrum of eigenstates and -values $\{\psi_i, E_i\}$. Eq.\,\ref{eq:CFT} can be rewritten as 
\begin{equation*}
S( \omega) \sim \frac{1}{Z} \sum\limits_{i} e^{-\frac{E_i}{kT}}  \frac{1}{\pi}
\text{Im} G_{i,i}(\omega),
\label{eq:CFT2}
\end{equation*}
so only a few lowest thermally populated states are actually needed, while the Green's function 
\begin{equation*}
G_{i,i}(\omega) = \bigg\langle \psi_i \bigg| \hat{T}^{\dagger} \frac{1}{\omega - \hat{H} +  i  \gamma/2} \hat{T} \bigg| \psi_i \bigg\rangle
\label{eq:transition}
\end{equation*}
can effectively be calculated using Lanczos tridiagonalization algorithm as implemented in the \texttt{Quanty} software used in our calculations \cite{Maurits_MLFT, retegan_crispy, Lanczos_1950, Pavarini-Koch-Coleman_Book}.

For $O_h$ symmetry, the crystal field is generated by the surrounding six oxygen ligands forming an octahedron, which leads to a splitting of the five $3d$ orbitals into three $t_{2g}$ and two $e_g$ orbitals. They are separated by an energy quantified by $10Dq$, which is typically between $1$ and $2\;\text{eV}$. Usually, $10Dq$ is lower than the Hund's rule energy $J_\text{H}$, \cite{Dagotto_Book, vdLK}, resulting in a high-spin (HS) configuration ${t_{2g}{\uparrow}}^3 {e_g{\uparrow}}^1$. 

The lattice mismatch with the STO substrate results in a tetragonal distortion, which lowers the symmetry to $D_{4h}$. This requires the introduction of two additional parameters $Ds$ and $Dt$ to quantify the lifting of the degeneracy of the $e_g$ and $t_{2g}$ orbitals: 
\begin{equation*}
\Deg = E_{d_{3z^2-r^2}} - E_{d_{x^2-y^2}} = - 4 Ds - 5 Dt,
\end{equation*}
\begin{equation*}
\Dttwog = E_{d_{xz,yz}} - E_{d_{xy}} = - 3Ds + 5 Dt.
\end{equation*}

Both splittings are mainly controlled by the neighboring oxygen anions in such a way that positive $\Deg$ and $\Dttwog$ correspond to an increase, and negative $\Deg$ and $\Dttwog$ correspond to a decrease of the in-plane bond-lengths with respect to the out-of-plane ones. As long as the tetragonal distortion is dominant, $\Deg$ and $\Dttwog$ should thus have the same sign.

Spin--orbit (SO) coupling is included for both the $2p$ and $3d$ shells, with SO coupling constants $\zeta_{2p}$ and $\zeta_{3d}$, respectively, where the former leads to the energy splitting of $3\zeta_{2p}/2$ between the  $L_2$ and $L_3$ peaks in the XAS spectrum.

The spherical part of the Coulomb repulsion between electrons in $3d$ orbitals is parameterized by $U_{dd}$. Once a $2p$ core hole is created during the absorption process, interactions between the $2p$ and $3d$ electrons have to be taken into account. The spherical part of this interaction is parameterized by $U_{pd}$. The ratio $U_{dd}/U_{pd} = 0.83$ follows from refs.~\onlinecite{Bocquet_1992_general} and~\onlinecite{Bocquet_1992_SFO, Andersen/Klose/Nohl, Fujimori/Minami}. The corresponding multipole interactions are parametrized by the Slater integrals $F_{dd}^2$, $F_{dd}^4$, $F_{pd}^2$ (direct exchange terms) and $G_{pd}^1$ and $G_{pd}^3$ (indirect exchange terms). Values for the atomic case, obtained numerically within Hartree--Fock theory, are provided in Ref.~\onlinecite{Maurits_thesis}. For solids, one typically has to scale all Slater integrals down by a factor of $\beta \lesssim 0.8$ to account for intra-atomic relaxation effects \cite{Thole_Lanthanides, Waddington_Dirac-Fock, vdLK, Lynch_Cowan}. Since relaxation can affect the $dd$ and $pd$ Slater integrals differently, we will use two distinct scaling factors, $\beta_{dd}$ and $\beta_{pd}$.

In MLFT, the ligand O $2p$ orbitals are explicitly taken into account. The spherical part of the difference between the Mn $3d$ on-site energies and the O $2p$ on-site energies is parameterized by $\Delta$ (charge-transfer energy in the Zaanen--Sawatzky--Allen scheme \cite{ZSA}). We neglect the impact of the small tetragonal distortion on the hopping and adopt the hopping strength in $O_h$ symmetry. Between the O $2p$ and the Mn $3d$ orbitals this is described by $V_{e_g} = -\sqrt{3} pd\sigma$ and $V_{t_{2g}} = 2 pd\pi$, with $pd\sigma/pd\pi = -2.0$ \cite{Bocquet_1992_general, Bocquet_1992_SFO, Andersen/Klose/Nohl, Fujimori/Minami}. The hopping among the O $2p$ orbitals is parametrized by a ligand orbital energy splitting $10Dq_\text{L}$  \cite{GHS_RNO_DC}.

Due to the explicit consideration of the ligand O $2p$ orbitals, charge-transfer configurations $d^{4+k} \underline{L}^k$ are included in the basis set, in addition to the purely ionic $d^{4} \underline{L}^0$: 
\begin{equation*}
\ket{\psi_\text{GS}} = \sum_{k=0}^6 c_ k\ket{d^{4+k} \underline{L}^k}
\end{equation*}
This allows the treatment of covalent compounds.

Finally, we also calculate the magnetic moment. In its general form, the magnetic moment operator reads $\boldsymbol{\mu}=- \mu_\text{B} (g_s \mathbf{\hat S} + \mathbf{\hat L})$, where we adopt a definition, in  which $\mathbf{\hat S}$ and $\mathbf{\hat L}$ are in units of $\hbar$. We want to ensure the comparability of our results with those of magnetometry measurements, which are performed in a magnetic field. The orientation of this field defines a convenient quantization axis, which we assume to be the $z$-axis. Therefore, rather than working with $\mathbf{\hat S}^2$ and $\mathbf{\hat L}^2$, we work with the $z$-projections $\hat S_z$ and $\hat L_z$. To obtain the maximal projection of the local magnetic moments, we introduce an auxiliary Weiss term to our Hamiltonian, which is sufficiently strong to fully saturate these moments at $T=5~\text{K}$. We then evaluate the expectation values $\langle\hat S_z\rangle$ and $\langle\hat L_z\rangle$, to obtain the magnetic moment $m_z$=$-\mu_\text{B} (g_s\langle\hat S_z\rangle+\langle\hat L_z\rangle)$.

We have performed our calculations at $5~\text{K}$ not only to ensure saturation, but also since magnetometry measurements are typically performed at low temperatures of this order of magnitude. Here, we make the reasonable assumption that the MLFT parameter values, which we need to evaluate the expectation values $\langle\hat S_z\rangle$ and $\langle\hat L_z\rangle$, even though obtained from fits to $300~\text{K}$ XAS data, are valid to a good approximation at low temperatures as well.

\section{Results}
\label{sec:Results}

\begin{figure}[tb]
\includegraphics[width = 1.0\columnwidth]{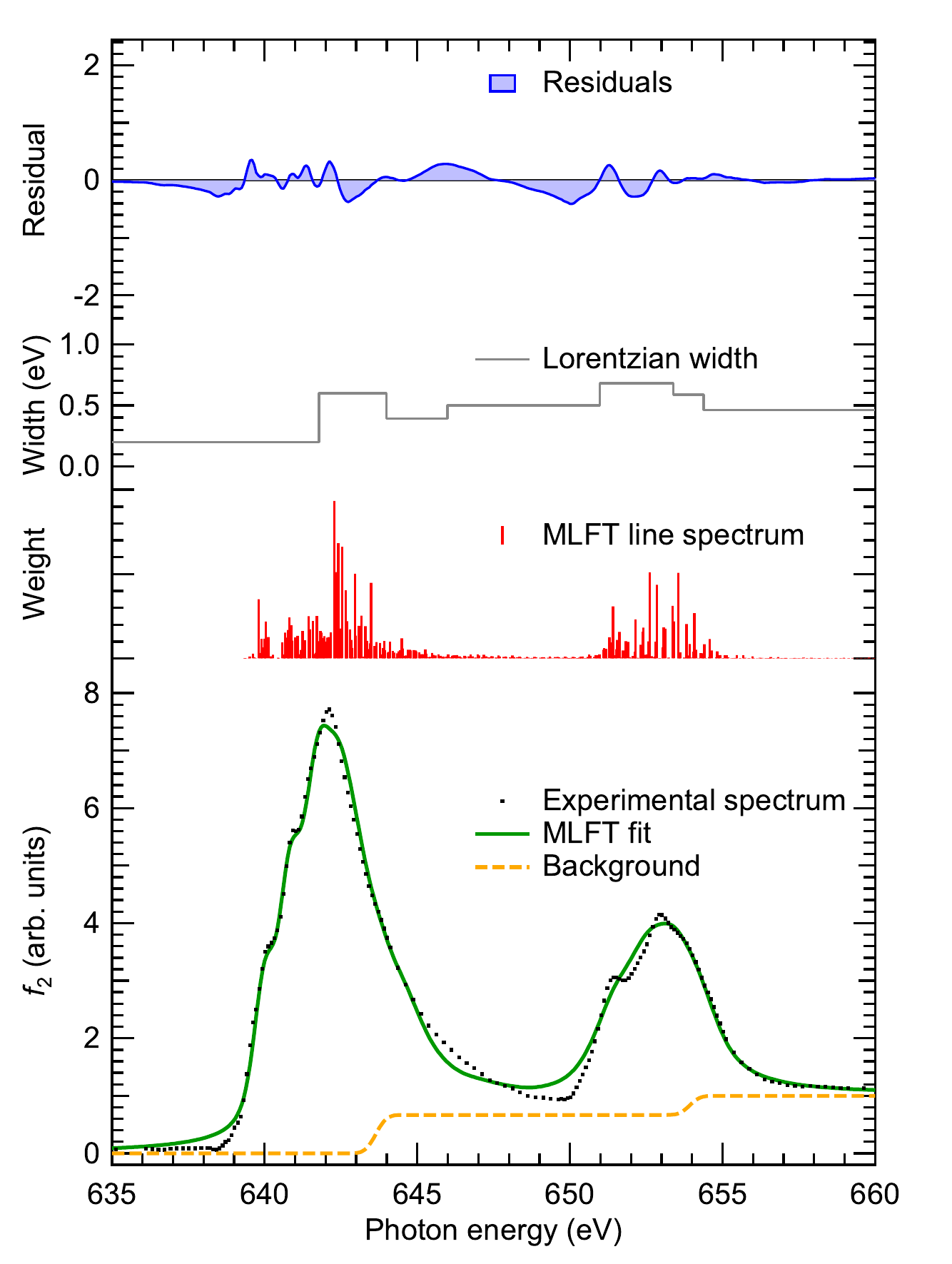}
\caption{(Color online) Experimental Mn$^{3+}$ $L_{2,3}$ X-ray absorption spectrum, together with the best fit obtained with our MLFT model. The model, as well as the Chantler background and the Lorentzian widths are described in the text. The values of the model parameters are shown in Table~\ref{tab:params-lft}. Red vertical bars denote the individual transitions from the initial- to the final-state multiplet, where the bar heights correspond to the relative weights of these transitions. }
\label{fig:layout1}
\end{figure}

We now discuss the fitting procedure of the measured spectrum with our MLFT model assuming $D_{4h}$ symmetry. Our initial analysis and optimization of all model parameters showed that it is sufficient to only refine $10Dq$, $\Deg$, $\Dttwog$, $\betadd$ and $\betapd$ in the final stage of the fit. In the following, we describe the handling of the remaining parameters not included in this final stage.

\begin{figure*}[tb]
	\includegraphics[width=\linewidth]{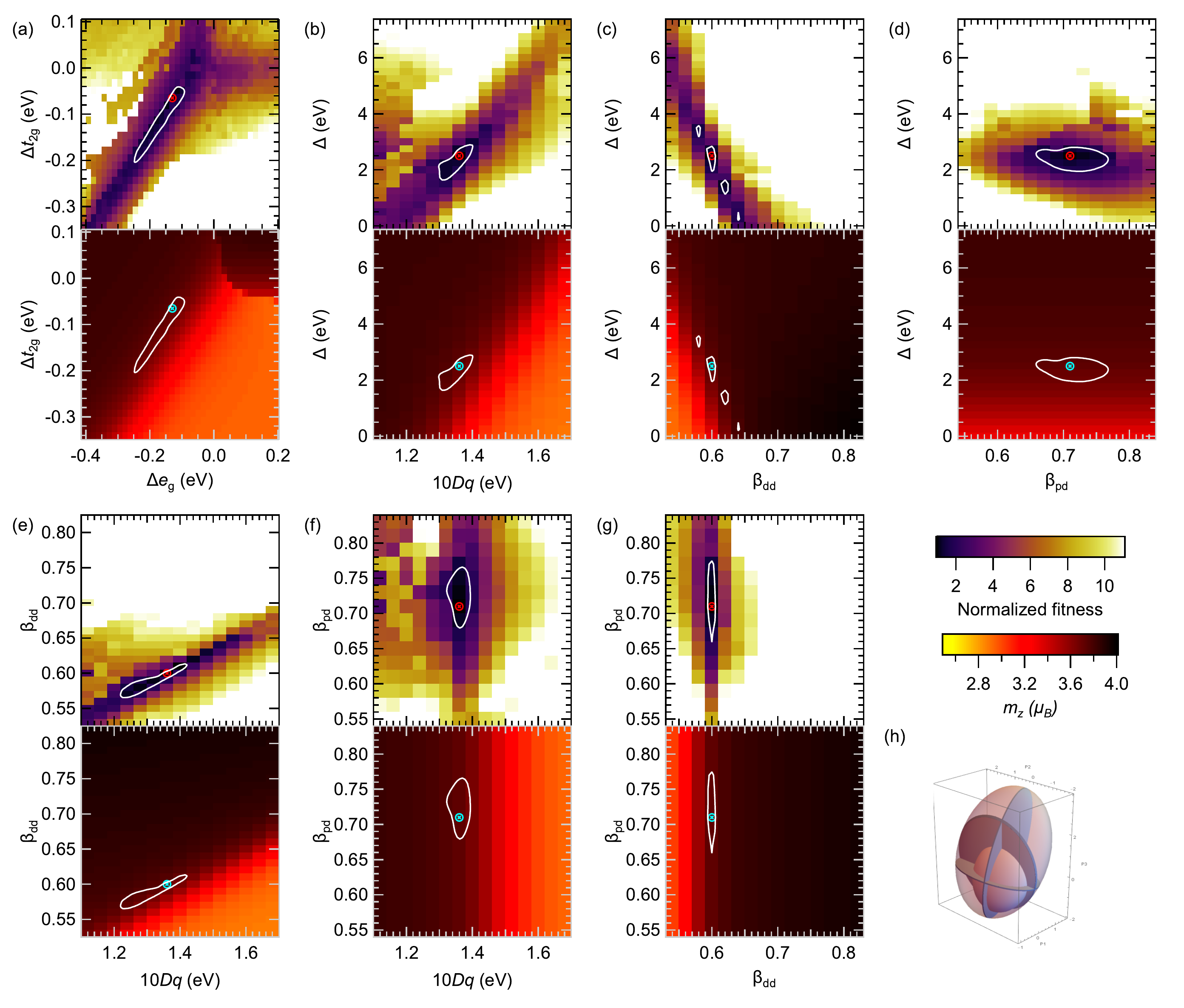}
	\caption{(Color online) Maps of the fitness function $f_{j,k}(x_j,x_k)$ [upper half of the panels (a) -- (g)], along with the corresponding magnetic moment projection of Mn,  $m_z$  [lower half of the panels]. The maps represent two-dimensional cuts through parameter space and are obtained by varying the values of the indicated parameter pairs $\{x_j, x_k\}$ around their optimized values $\{\xi_j, \xi_k\}$ resulting from the fit, while keeping all other parameters fixed. The fitness function $f_{j,k}(x_j,x_k)$ is normalized to 1 at its minimum, which is marked with a cross. The white contours surround the confidence areas with $f_{j,k}(x_j,x_k)\leq 2$. The used parameter pairs are as follows: (a) $\Deg$ and $\Dttwog$, (b) $10Dq$ and $\Delta$, (c) $\betadd$ and $\Delta$, (d) $\betapd$ and $\Delta$, (e) $10Dq$ and $\betadd$, (f) $10Dq$ and $\betapd$, (g) $\betadd$ and $\betapd$. Panel (h) illustrates how the shown two-dimensional cuts and confidence areas relate to the multidimensional parameter space and confidence region, which is assumed ellipsoidal for the sake of simplified representation.}
	\label{fig:fitd_fitmommaps}
\end{figure*}

The charge-transfer energy $\Delta$ and the hopping parameters $V_{e_g}$ and $V_{t_{2g}}$ were varied within reasonable boundaries in the first stages of the fit, until a good agreement and stable behavior had been achieved. The values resulting from this approach are compatible with those reported in refs.~\onlinecite{Saitoh_LSMO_95, Saitoh_LXO_95, Taguchi/Altarelli, Cho_2009, Chainani, Zampieri, Park_FMtr}, hence they were subsequently fixed for the final fits. Due to its importance for the $d$-level filling $n_d$, we have included $\Delta$ in the list of parameters, whose values we later scrutinize when discussing the robustness of our entire fit (Fig. \ref{fig:fitd_fitmommaps}). The same approach was taken for $10Dq_\text{L}$, which turns out to only have a marginal influence on the spectrum.

\begin{figure}[tb]
	\centering
	\includegraphics[width = 1.0\columnwidth]{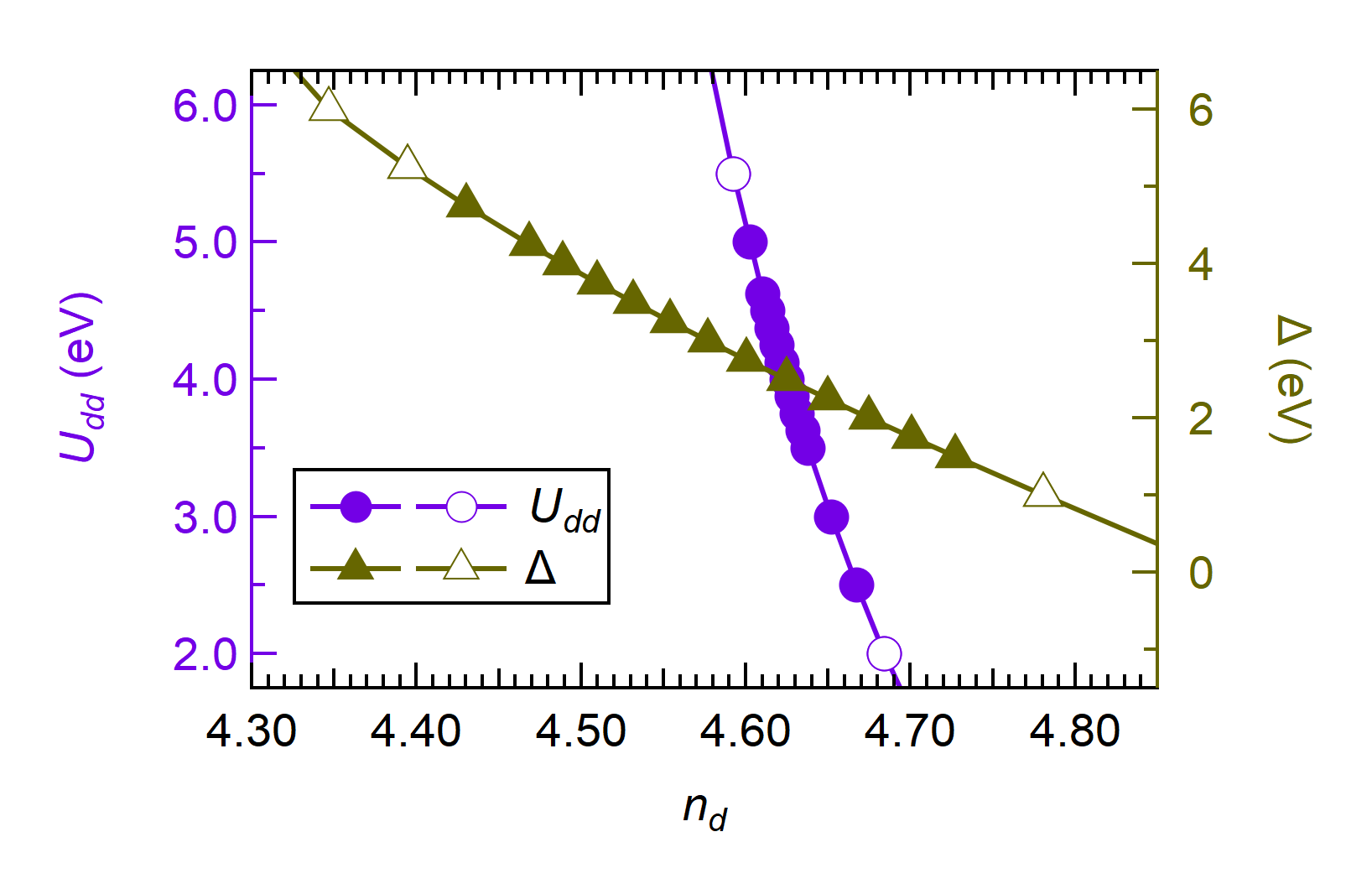}
	\caption{(Color online) Variation of the $3d$ filling as function of $U_{dd}$ (circles) or $\Delta$ (triangles) around their best-fit values shown in table~\ref{tab:params-lft}. Closed (open) symbols denote $U_{dd}$ and $\Delta$ values within (beyond) the reasonable boundaries $2.5\;\text{eV} \leq U_{dd} \leq 5\;\text{eV}$ and $1.5\;\text{eV} \leq \Delta \leq 4.75\;\text{eV}$.}
	\label{fig:ndtrafo}
\end{figure}

\begin{figure}[tb]
	\centering
	\includegraphics[width = 1.0\columnwidth]{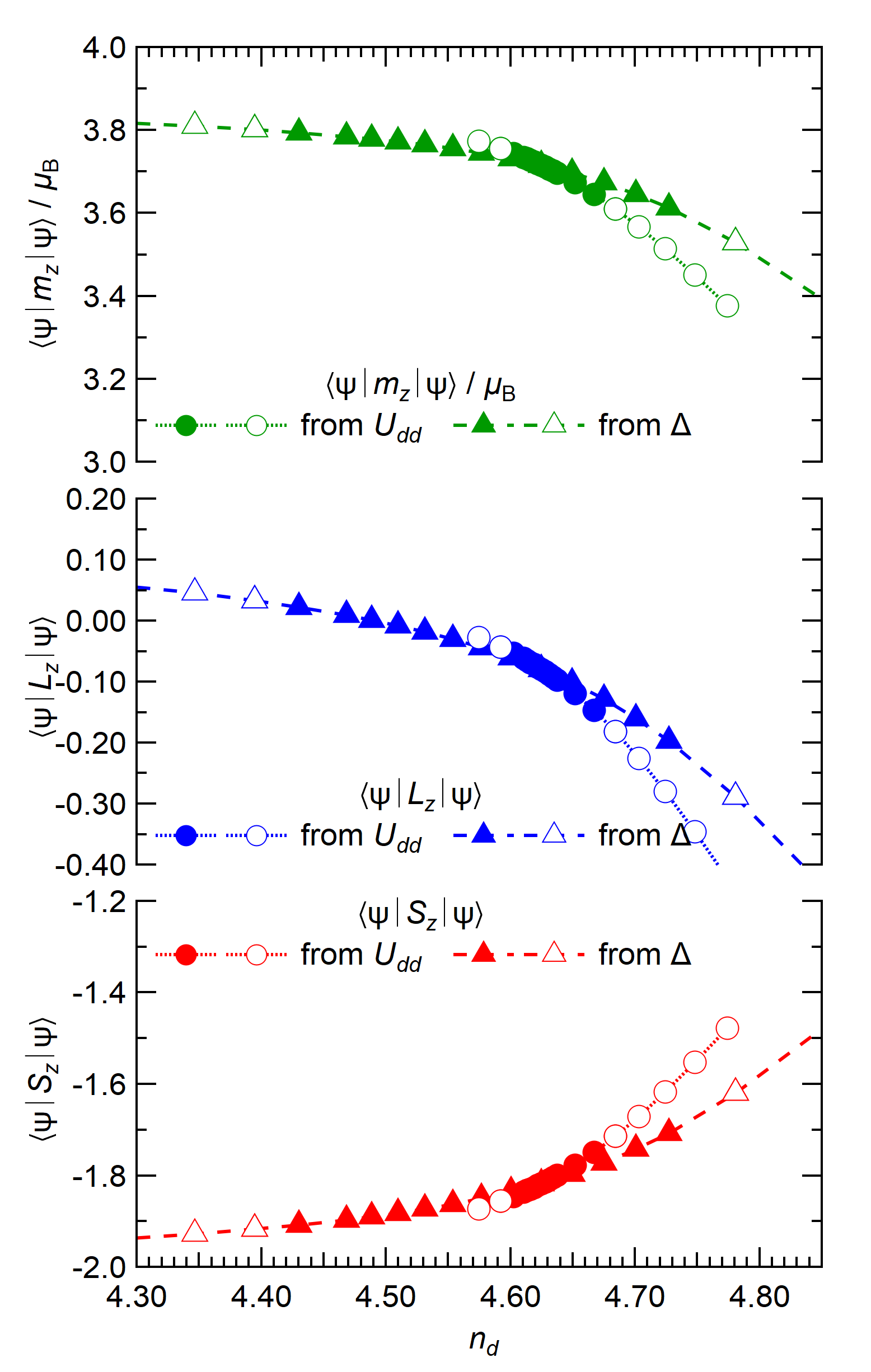}
	\caption{(Color online) Magnetic, spin and orbital moments, $m_z$, $\langle\hat S_z\rangle$, $\langle\hat L_z\rangle$, respectively, as a function of the $3d$ filling, which is controlled by either $U_{dd}$ (circles) or $\Delta$ (triangles). Closed (open) symbols correspond to $U_{dd}$ and $\Delta$ values within (beyond) reasonable boundaries.}
	\label{fig:ndmom}
\end{figure}

Furthermore, the values of $\zeta_{2p}$ and $\zeta_{3d}$ are known with such precision \cite{Maurits_thesis}, that within the remaining uncertainty margin, their variation turned out not to alter the fit in a significant way. This, and the substantial increase in computational time to consider two additional parameters, justifies to fix their values \cite{Maurits_thesis}.  As we demonstrate in Figs.~\ref{fig:ndtrafo} and~\ref{fig:ndmom}, the variation of $U_{dd}$ within a physically meaningful range has a very small impact on key results such as the $3d$ level filling and the magnetic moment. Therefore, we fix $U_{dd}$ (and $U_{pd}$, where similar considerations can be employed) according to refs.~\onlinecite{Taguchi/Altarelli, Cho_2009, Chainani, Zampieri, Park_FMtr}.

Finally, our model accounts for Lorentzian core-hole lifetime broadening $\gamma(E)$, which is described by the step function shown in Fig.~\ref{fig:layout1}. The step heights and the most crucial step positions were included in the final fit. Appropriate restrictions were applied to ensure that $\gamma(E)$ stays within the reasonable range $[0.2\;\text{eV}, 0.8\;\text{eV}]$, and that it increases largely monotonically within this range. To model the effects of the experimental resolution \cite{Taguchi/Altarelli, Volodya_SmB6, Abdul_VvCr}, the life-time broadened curve was additionally convolved with a Gaussian, whose FWHM was also treated as a fit parameter.

After having disencumbered our model as described, we performed the final fit to the experimental data using a genetic fit algorithm \cite{Sebastian_2014, Refl_Genetic_Algorithm}. We define the fitness function 

\begin{gather}
 \nonumber f(x_1, x_2, ... ,x_n) =  \\ 
\sum_{m=1}^M \left[ S_\text{MLFT}(x_1, x_2, ..., x_n, \omega_m) - S_\text{XAS}(\omega_m)\right]^2,
\label{eq:fitnessfunction}
\end{gather}

where $x_i$ is a fitting parameter, $S_\text{XAS}(\omega)$ is the experimental spectrum, $S_\text{MLFT}(x_1, x_2, ..., x_n, \omega)$ is the calculated MLFT spectrum, and the sum runs over all measured energy points.

The lower part of Fig.~\ref{fig:layout1} shows a comparison of the optimal theoretical curve resulting from the fit with the experimental data. Above that, we show the original, properly weighted but unbroadened multiplet peaks, the life-time broadening $\gamma(E)$ and the residuals. The values of the corresponding fitting parameters are listed in tables~\ref{tab:params-lft} and~\ref{tab:finalfitslater}.

\begin{table}[t]
	\renewcommand*{\arraystretch}{1.4}
	\begin{tabular*}{.22\textwidth}[t]{l @{\extracolsep{\fill}} c}
		\toprule
		Fit parameters & Results \\
		\midrule
		$10Dq\;\left(\text{eV}\right)$ & $1.36$ \\
		$\Delta e_\text{g}\;\left(\text{eV}\right)$ & $-0.13$ \\
		$\Delta t_\text{2g}\;\left(\text{eV}\right)$ & $-0.065$ \\
		$\betadd$ & $0.60$ \\
		$\betapd$ & $0.71$ \\
		$\text{Gauss FWHM}\;\left(\text{eV}\right)$ & $0.61$ \\
		\bottomrule
	\end{tabular*}
	\begin{tabular*}{.25\textwidth}[t]{l @{\extracolsep{\fill}} c}
		\toprule
		Early fit parameters & Results \\
		\midrule
		$\Delta\;\left(\text{eV}\right)$ & $2.5$ \\
		$V_{e_g}\;\left(\text{eV}\right)$ & $2.50$ \\
		$V_{t_{2g}}\;\left(\text{eV}\right)$ &  $1.44$ \\
		\midrule
		Fixed parameters & Results \\
		\midrule
		$10Dq_\text{L}\;\left(\text{eV}\right)$ & $0.8$\\
		$U_{dd}\;\left(\text{eV}\right)$ & $4.0$ \\
		$U_{pd}\;\left(\text{eV}\right)$ & $4.8$ \\
		\bottomrule
	\end{tabular*}\\
	\caption{Parameters for the MLFT fit shown in Fig.~\ref{fig:layout1}.}
	\label{tab:params-lft}
\end{table}

\begin{table}[b]
	\renewcommand*{\arraystretch}{1.4}
	\begin{tabular*}{.5\textwidth}{l @{\extracolsep{\fill}} cc}
		\toprule
		& Initial state ($2p^6$ $3d^4$) & Final state ($2p^5$ $3d^5$) \\ \midrule
		$F^2_{dd}$ & $6.85$ & $7.33$ \\
		$F^4_{dd}$ & $4.29$ & $4.59$ \\
		$F^2_{pd}$ & & $4.96$ \\
		$G^1_{pd}$ & & $3.68$ \\
		$G^3_{pd}$ & & $2.09$ \\ \bottomrule
	\end{tabular*}
	\caption{Slater integrals for the MLFT fit shown in Fig.~\ref{fig:layout1}. All values are given in eV.}
	\label{tab:finalfitslater}
\end{table}

Besides the nominal ionic, zero--ligand-hole configuration $d^{4} \underline{L}^0$, our fit reveals significant contributions to the ground state wavefunction from the one-- and two--ligand-hole configurations. These three configurations $d^{4} \underline{L}^0$, $d^{5} \underline{L}^1$ and $d^{6} \underline{L}^2$ account for $40\%$, $47\%$ and $12\%$, respectively. Other configurations were neglected owing  to their vanishing contribution. A test calculation allowing up to 6 ligand holes showed that the total contribution of these neglected configurations does not exceed $3\%$ for a wide range of crystal fields and scalings of the Slater integrals. The strong contribution of the $d^{5} \underline{L}^1$ configuration can be understood as being due to the stabilized half full shell.

Hence, the first significant result of our investigation is the strongly covalent character of LSMO: There is a considerable spread of the ground state over several configurations, which stresses the importance of using ligand field theory. In particular, it is insufficient to only discuss the nominal valence (in our case $3+$) when dealing with manganites. A calculation of physical quantities such as the magnetic moment under the assumption of the nominal $d$-shell occupation and disregarding charge transfer would obviously lead to erroneous results. 

A second important result we have obtained is the energy splittings of the $e_g$ and $t_{2g}$ levels, $\Deg = -0.13\;\text{eV}$ and $\Dttwog = -0.065\;\text{eV}$. The negative sign indicates that the $d_{3z^2-r^2}$ orbital has a lower energy than the $d_{x^2-y^2}$ orbital, and the $d_{xz,yz}$ orbitals have a lower energy than the $d_{xy}$ orbital. We point out that by using linearly ($\sigma$) polarized light, we are highly sensitive to the sign of the energy splittings due to the pronounced orientational dependence of the corresponding orbitals, and due to the fact that their occupation partially changes upon changing the sign of the splitting. This sensitivity is corroborated by the fitness analysis presented later (see Fig.  \ref{fig:fitd_fitmommaps}), and by simulations demonstrating a fundamental modification of the spectral shape upon a sign change (not shown here). 

How are these energy splittings related to the strain the substrate exerts on the film? There are three conceivable mechanisms, how strain can affect the LSMO unit cell: first, it can lead to rotations and tilts of the MnO$_6$ octahedra; second, it can deform the octahedra; and third, it can provide a preferential orientation to establish a static Jahn--Teller effect.

The magnitude of the $3d$ level splittings is remarkable in view of the very small lattice mismatch of our film with the STO substrate ($\lesssim 0.2\%$) and the fact that the three Mn--O distances are nearly identical in the bulk at this doping level \cite{Kawano_LSMO}. Whereas LSMO is prone to tilts and rotations of the MnO$_6$ octahedra, we do not expect them to result in significant $3d$ level splittings in the absence of a simultaneous modification of the Mn--O distances. A deformation of the octahedra changing these distances would naturally impact the $3d$ levels. However, in compounds without relevant Jahn--Teller physics, $e_g$ splittings of the size we observe require a lattice mismatch of well above $1\%$ \cite{Rogge_CFO, Wu_Nickelates}: In LaNiO$_3$ films, for instance, the $e_g$ splitting scales monotonically and roughly linearly with $m$, ranging from about $-0.1\;\text{eV}$ for $m=-2\%$ to about $0.3\;\text{eV}$ for $m=+3.5\%$. Compared to this relation, the mismatch effect on  $\Deg$ in our sample is an order of magnitude larger. 

These considerations provide strong support for the third mechanism we propose above---namely the stabilization of a static Jahn--Teller effect. As described in section~\ref{sec:Experiment}, in the bulk La$_{7/8}$Sr$_{1/8}$MnO$_3$ would be in the dynamic Jahn--Teller O$^*$-phase, which is characterized by random distortions of the MnO$_6$ octahedra. However, at room temperature, at which our measurements were performed, it would be very close to the transition to a statically ordered Jahn--Teller phase ($T_{\text{JT}}\approx 280\;\text{K}$) and thus strongly susceptible to lattice strain. Our finding of disproportionately strong $3d$ level splittings thus suggests that lattice strain provides a preferential orientation for a spontaneous, collective Jahn--Teller deformation of the octahedra, characterized by an elongation of the out-of-plane Mn--O bonds relative to the in-plane ones. 

The identification of the Jahn--Teller effect as the underlying mechanism for the observed $3d$ splittings is supported by the fact that the magnitude of the $e_g$ splitting of $0.13\;\text{eV}$ we observe lies well within the range of Jahn--Teller energies reported in literature for the vicinity of our doping level, namely between $0.1\;\text{eV}$ and $0.2\;\text{eV}$ \cite{Tang_simu, Bhattacharya_JPCM}.

The values reported in Tables~\ref{tab:params-lft} and~\ref{tab:finalfitslater} based on our MLFT fit allow us to evaluate the expectation values of the spin and orbital angular momentum operators, $\langle\hat S_z\rangle$ and $\langle\hat L_z\rangle$, as described at the end of section~\ref{sec:Theory}, and to calculate the local magnetic moment $m_z$. We obtain $3.72\;\mu_\text{B}/\text{Mn atom}$, which is a further noteworthy result of our work. This value confirms that the Hund's rule energy $J_\text{H}$ is larger than the crystal field energy \cite{Dagotto_Book, vdLK} as parameterized by $10Dq$, $\Deg$, and $\Dttwog$, and is in line with many results obtained for La$_{1-x}$Sr$_x$MnO$_3$ with various techniques within a doping range of $0\leq x <0.3$  \cite{Jonker-van-Santen, Kotani_Magnetic_bubbles, Hu_LSMO/LAO, Monsen_LSMO, Wahler_LSMO_nanostructures, Borges_LSMO/MgO}. Notably, other authors report significantly lower moments, in many cases well below $2\;\mu_\text{B}$ \cite{Niu, Kim_0.12_LSMO, Hu_LSMO/LAO, Shibata_2018}.

Various reasons are possible for this spread of reported magnetic moments. First, depending on the preparation process, the sample composition quality might vary, and imperfections such as ferromagnetically ``dead'' layers or domains might occur \cite{Kodama_Dead_Layer, Monsen_LSMO, Shibata_2018, Borges_LSMO/MgO, Huijben_LSMO}, which reduce the total ferromagnetic volume and therefore distort the normalization of magnetometry data, as described below. The nature of dead layers is not entirely clear, and also not generalizable to different systems \cite{Huijben_LSMO, Peng_Dead_Layer}. However, the most obvious realization would be a degradation of the surface layer due to \emph{ex-situ} conditions, for instance due to a reduction of the Mn valence. 

Second, given the complexity of the phase diagram, the magnetic properties can critically depend on the precise doping level, which might be difficult to control during growth. Hence, a precise experimental characterization of not only the magnetic but also of the electronic properties is required. 

Regarding the magnetic characterization, considerable confusion arises from the fact that in some publications the \emph{ordered} magnetic moment is reported, whereas in others the \emph{local} one, irrespective of its alignment relative to its neighbors. This is owed to the sensitivity of the different employed techniques.

Some techniques such as (SQUID) magnetometry, are more sensitive to the ordered moment, since they measure the susceptibility of the entire sample, which is much larger in the FM phase as compared to the paramagnetic phase. In order to obtain the ordered moment in nominally ferromagnetic samples using magnetometry, one needs to correct for inhomogeneities and magnetically ``dead'' layers or domains, which can be cumbersome. If, on the other hand, one is interested in the local moment, one needs to magnetically saturate the entire sample, including the ``dead'', presumably paramagnetic domains. However, saturation of such paramagnetic domains requires large fields, which are often not being explored, or which are even inaccessible, in a magnetometry measurement.

Other techniques, such as resonant X-ray spectroscopy, can very precisely probe the full local moments, irrespective of whether or not they are in an ordered phase. In this work we use X-ray absorption spectroscopy, which allows us a characterization of both the local electronic and magnetic properties: As described in section~\ref{sec:Theory}, we first determine the local electronic configuration, from which we then obtain the local magnetic moment $m_z$. This local character of XAS should be kept in mind, when comparing our result to published magnetometry and related results.

Depending on the theoretical approach, calculated moments often also deviate from the experimental results. An obvious discrepancy would arise if Hund's rules were applied to the nominal Mn valence, in our case $3+$, since this would not take into account charge transfer from the oxygen ligands and the resulting admixtures of the $d^{5} \underline{L}^1$ and $d^{6} \underline{L}^2$ configurations to the nominal $d^{4} \underline{L}^0$.

\begin{figure}[tb]
	\centering
	\includegraphics[width = 1.0\columnwidth]{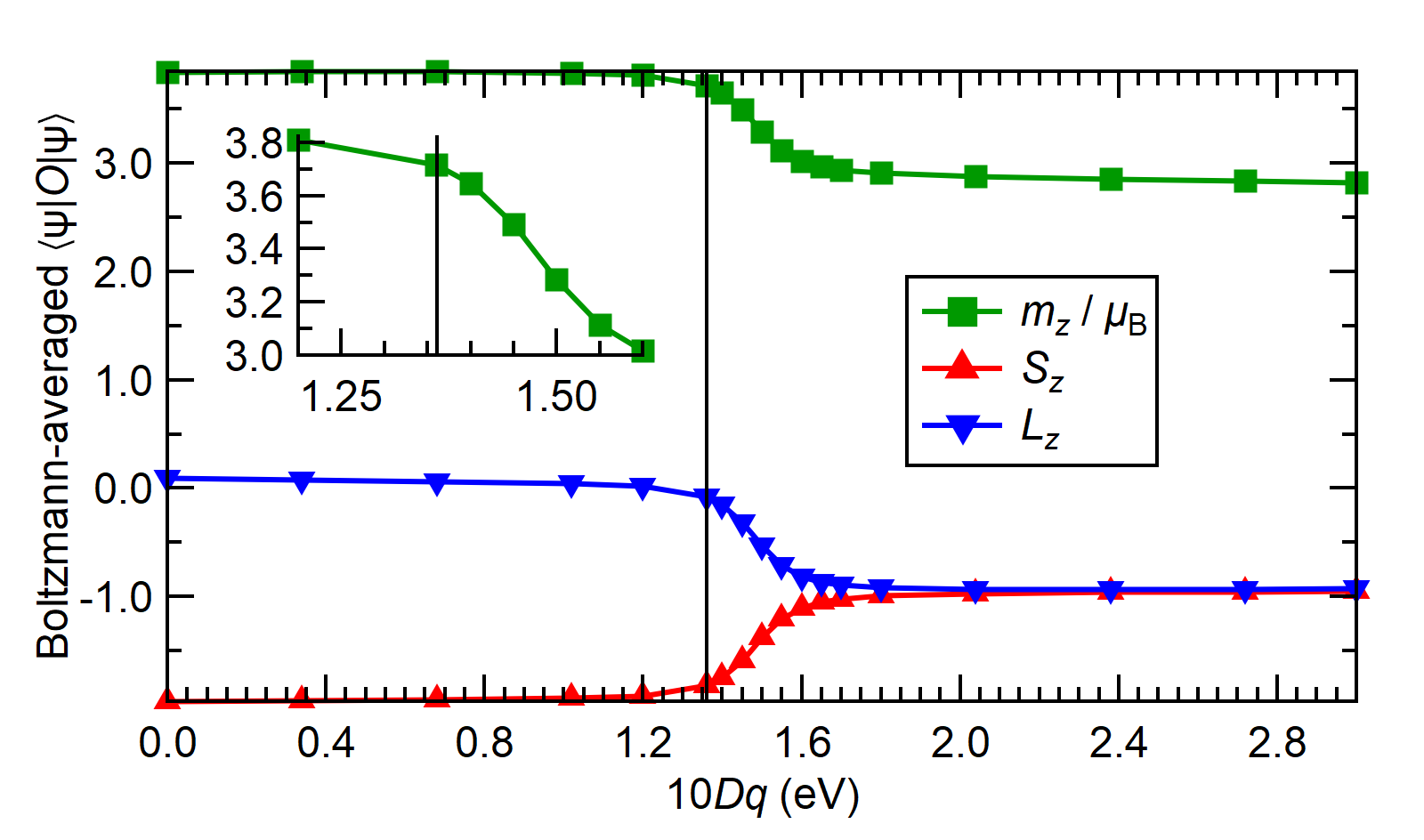}
	\caption{(Color online) Magnetic, spin and orbital moments, $m_z/\mu_\text{B}$, $\langle\hat S_z\rangle$ and $\langle\hat L_z\rangle$, respectively, as a function of the strength of the local crystal field, parametrized by $10Dq$. The value $10Dq = 1.36\;\text{eV}$, obtained from our best fit (see table \ref{tab:params-lft}), is marked by a vertical line. Inset: zoom-in to $m_z/\mu_\text{B}$ in the vicinity of $10Dq= 1.36\;\text{eV}$.}
	\label{fig:dqmom}
\end{figure}

Further modifications of the moment can be induced by crystal field effects. Usually, the Hund's rule energy is assumed to be substantially larger than the crystal field splitting $10Dq$ \cite{Fazekas_Book, de-Groot-Kotani_Book, Dagotto_Book, vdLK}, resulting in a HS state for Mn$^{3+}$. In Fig.~\ref{fig:dqmom}, we show the effect of variations of $10Dq$ on $\langle \hat{S}_z \rangle$, $\langle \hat{L}_z \rangle$ and the magnetic moment $m_z$. Obviously, whereas Mn$^{3+}$ is in the HS state, it is on the verge of a transition to a low-spin (LS) state: The magnetic moment would be reduced below $3.0\;\mu_\text{B}/\text{Mn atom}$ by an additional crystal field splitting of only $150\;\text{meV}$. This is well within the range achievable by tetragonal distortions, which can occur due to substrate strain, but conceivably also due to interfacial reconstructions or compositional or structural degradations at the surface or in inhomogeneous samples. Hence, the vicinity of Mn$^{3+}$ to a spin state transition might further contribute to the spread of reported magnetic moments.

Another quantity having an impact on the moment is the $3d$ level filling $n_d$, which can easily change due to unintentional doping caused by impurities and defects or surface degradation. Doping can also occur due to oxygen vacancies, the propensity to whose formation strongly depends on the Sr content \cite{Pavone_first_principles}, and possibly also on the surface quality and the growth and annealing conditions.

The $3d$ level filling is not an MLFT input parameter, which can be freely varied. Rather, it results from the interplay of other parameters, mainly $U_{dd}$ and $\Delta$, and can be obtained as a corresponding expectation value. Based on our final fit, we obtain $n_d \approx 4.625\;\text{electrons}$. 

Therefore, to investigate the impact on $n_d$, we have first varied $U_{dd}$ and $\Delta$ within reasonable boundaries.   In Fig.~\ref{fig:ndtrafo} we report how this variation changes $n_d$. As can be seen, the $3d$ filling depends only weakly on $U_{dd}$. However, it can be modified by about $\pm 0.2\;\text{electrons}$ by varying $\Delta$.

Hence, whereas we do not have direct control of the doping within our theoretical approach, varying $U_{dd}$ and $\Delta$ permits a rough estimate of the impact of unintentional doping, covering a substantial range of the phase diagram. In Fig.~\ref{fig:ndmom}, we show how $\langle \hat{S}_z \rangle$, $\langle \hat{L}_z \rangle$ and $m_z$ change as a result of the modified filling. The magnetic moment remains above $3.5\;\mu_\text{B}/\text{Mn atom}$ in all cases. This suggests that unintentional doping cannot be a major factor in reducing the moment to values as low as $2\;\mu_\text{B}/\text{Mn atom}$ or less, which are values reported in some publications \cite{Niu, Kim_0.12_LSMO, Hu_LSMO/LAO}. Fig.~\ref{fig:ndmom} also shows that $U_{dd}$ and $\Delta$ variations have a somewhat different impact on $m_z$. This is due to the obvious fact that $U_{dd}$ and $\Delta$ control further physical properties of the system beyond the $3d$ filling.

Finally, we discuss the robustness of our fits and of the resulting magnetic moment. To this end, we consider two-dimensional cuts through parameter space, in which two parameters $x_j$ and $x_k$ are varied, while the remaining parameters $\{x_i\}_{i\neq j, i\neq k}$ are fixed to their optimal values $\{\xi_i\}_{i\neq j, i\neq k}$ (Fig. \ref{fig:fitd_fitmommaps}). 

It is convenient to normalize the fitness function $f(x_1, x_2, ..., x_n)$ defined in Eq.\,\ref{eq:fitnessfunction} to 1 at its minimum $f(\xi_1, \xi_2,..., \xi_n)$, which was established in the fit. Also, to simplify our notation, we define the partial fitness function $f_{j,k}(x_j,x_k)$ as $f(\xi_1,..., \xi_{j-1}, x_j, \xi_{j+1},...,\xi_{k-1}, x_k, \xi_{k+1},...,\xi_n)$. The upper half of each panel shows this partial fitness function $f_{j,k}(x_j,x_k)$ around its minimum $f_{j,k}(\xi_j,\xi_k)=1$ at $(\xi_j,\xi_k)$. The white contours indicate confidence areas, within which $f_{j,k}(x_j,x_k)\leq2$. The sizes of the maps were chosen to generously encompass these confidence areas.

In panels (b) through (g), we examine pairs of the parameters $\Delta$, $10Dq$, $\betadd$ and $\betapd$, which are all nonzero in $O_h$ symmetry. In panel (a), we show a $\Deg$-$\Dttwog$ map---both these parameters are relevant when the symmetry is further broken to $D_{4h}$. Obviously, different parameter pairs are correlated to different degrees: For instance, decreasing the value of $\Deg$ within certain boundaries can be compensated by simultaneously decreasing the value of $\Dttwog$, to remain within the $2 f_{j,k}(\xi_j,\xi_k)$ confidence area. On the other hand, $10Dq$ and $\betapd$ appear largely uncorrelated.

The fitness maps demonstrate the quality of our fits, and allow us to discuss the correlations and the fit errors. In panel (a), the asymmetric location of the minimum within the confidence area indicates rather relaxed lower boundaries but much stricter upper boundaries for $\Deg$ and $\Dttwog$: Their values could be up to $\approx 0.1\;\text{eV}$ lower but only up to $\approx 0.02\;\text{eV}$ higher than the values established in the fit. Obviously, such considerations strongly depend on the required level of confidence, which is to some degree arbitrary. We have chosen to regard the $200\%$ contours, which allows a rather fair estimate of the fit errors. A closer scrutiny of all maps suggests $10Dq$ to lie in the interval between $1.2$ and $1.4\;\text{eV}$, $\Delta$ between $1.75$ and $3\;\text{eV}$, $\betadd$ between $0.57$ and $0.62$ and $\betapd$ between $0.65$ and $0.77$. Irrespective of the specific choice of the level of confidence, such estimated intervals should be used with caution, since they provide only a very limited information about the complex, irregularly shaped confidence region and the parameter correlations it describes. 

Now that we have established the robustness of our fit, we consider the implications for the resulting magnetic moment projection of Mn, $m_z$.  In the lower halves of the panels in Fig. \ref{fig:fitd_fitmommaps} we show how $m_z$ depends on the same parameter pairs as in our discussion of the fitness maps. In all cases, $m_z$ remains above $3.5\;\mu_{\text{B}}$ within the $2 f_{j,k}(\xi_j,\xi_k)$ confidence area, lending further support to the robustness of the established magnetic moment. In particular, despite its proximity to the spin state transition (see Fig. \ref{fig:dqmom}), the system is manifestly on the high-spin side.

\section{Summary}
\label{sec:Summary}

In summary, we have combined X-ray absorption spectroscopy with multiplet ligand field theory to study the local electronic and magnetic properties of Mn in La$_{1-x}$Sr$_x$MnO$_3$ at $x=1/8$. Our sample is a $6\;\text{unit cells}$ thin film grown on a SrTiO$_3$ substrate. There is a small lattice mismatch of $\lesssim 0.2\%$ between film and substrate, resulting in compressive strain on the film. 

Our MLFT fit to the XAS data reveals a strongly covalent character, with the ionic configuration $d^{4} \underline{L}^0$ accounting for only $40\%$, and the one- and two-ligand-hole configurations contributing $47\%$ and $12\%$, respectively. 

The $e_g$ and $t_{2g}$ levels are split by $\Deg = -0.13\;\text{eV}$ and $\Dttwog = -0.065\;\text{eV}$, respectively, i.e. the $d_{3z^2-r^2}$ orbital has a lower energy than the $d_{x^2-y^2}$ orbital and the $d_{xz,yz}$ orbitals have lower energies than the $d_{xy}$ orbital. In the absence of Jahn--Teller fluctuations the present substrate strain of less than $0.2\%$ would be too small to explain these splittings. At $x=1/8$ doping, bulk La$_{1-x}$Sr$_x$MnO$_3$ is in the dynamic Jahn--Teller O$^*$ phase. Therefore, we interpret the disproportionately large $3d$ level splittings as resulting from a scenario, in which the compressive strain provides a preferential orientation for the MnO$_6$ octahedra to collectively elongate along the $c$ axis. The $e_g$ splitting of $\Deg = -0.13\;\text{eV}$ agrees well with the Jahn--Teller energies reported in the literature for the vicinity of our doping level, lending further support for our interpretation.

Based on the electronic structure and level occupations resulting from our MLFT fit, we obtain a local magnetic moment $m_z$ of $3.7\;\mu_{\text{B}}/\text{Mn}$, in agreement with a high-spin configuration, in which the Hund's rule energy $J_{\text{H}}$ is larger than the crystal field energy. A closer investigation shows that our system is nevertheless close to a HS-LS transition, which could be triggered by crystal field effects. On the other hand, we demonstrate that changes in the $3d$ level fillings of the order of magnitude which can occur by unintentional doping would not suppress the magnetic moment below $3.5\;\mu_{\text{B}}/\text{Mn}$.

Due to the proximity of the HS-LS transition, we have put particular emphasis on evaluating the robustness of our fits and in particular on the obtained magnetic moment. We show that with high confidence, the errors of the obtained values of the fitted MLFT parameters are sufficiently small as to warrant a magnetic moment of at least $m_z = 3.5\;\mu_{\text{B}}/\text{Mn}$.

The seeming discrepancy between the large value of $m_z$ we obtain and some values reported in the literature, which are in many cases below $2\;\mu_\text{B}$ \cite{Niu, Kim_0.12_LSMO, Hu_LSMO/LAO, Shibata_2018}, can be largely explained by the fact that different quantities are measured by different methods, even though the same terminology is used: With XAS, we determine the \emph{local} magnetic moment, whereas the widely used magnetometry methods determine the total magnetic moment of the sample, which is then normalized by the number of atoms assumed to be carrying a local moment. In particular for ferromagnetic samples, which contain ``dead'' regions or layers, or for samples which are not fully magnetically saturated, this normalization does not produce the same local moment as local spectroscopies yield.

\section*{Acknowledgments}

This work was funded by the Deutsche Forschungsgemeinschaft (DFG, German Research Foundation) Project-ID 258499086--SFB 1170 (projects C04, C06, and Z03), and the Natural Sciences and Engineering Research Council of Canada (NSERC). We thankfully acknowledge Maurits W. Haverkort for fruitful discussions. Part of research described in this paper was performed at the Canadian Light Source, a national research facility of University of Saskatchewan, which is supported by the Canada Foundation of Innovation (CFI), the Natural Sciences and Engineering Research Council (NSERC), the National Research Council (NRC), the Canadian Institutes of Health Research (CIHR), the Government of Saskatchewan, and the University of Saskatchewan.

\FloatBarrier

\nocite{*}

\bibliographystyle{apsrev4-2}


%

\end{document}